\documentclass[
reprint,
superscriptaddress,
amsmath,amssymb,
aps,
pra,
floatfix,longbibliography
]{revtex4-2}
\usepackage{graphicx}
\usepackage{dcolumn}
\usepackage[hypertexnames, breaklinks=true, bookmarksnumbered=true, bookmarksopen=true, colorlinks=true, linktocpage=true, citecolor=magenta, urlcolor=magenta, linkcolor=magenta]{hyperref}
\usepackage{amsmath,amssymb,amsfonts,amsthm}
\usepackage{mathtools} 
\usepackage[utf8]{inputenc}
\usepackage{txfonts}
\usepackage{bbm} 
\usepackage{bm} 
\usepackage{mathrsfs} 
\usepackage{xcolor}
\usepackage{float}

\allowdisplaybreaks

\theoremstyle{plain}

\newtheorem{proposition}{Proposition}
\theoremstyle{definition}

\theoremstyle{remark}

\usepackage{braket}
\newcommand{\ketbra}[2]{|#1\rangle\!\langle#2|}

\newcommand{\id}{\mathbbm{1}} 
\newcommand{\tr}{\mathrm{Tr}} 
\newcommand{\C}{\mathcal{C}} 

\begin{document}

\title{Preserving quantum correlations and coherence with non-Markovianity}
\author{Marek Miller}
\thanks{These authors contributed equally to this work.}
\affiliation{Centre for Quantum Optical Technologies, Centre of New Technologies, University of Warsaw, Banacha 2c, 02-097 Warsaw, Poland}
\author{Kang-Da Wu}
\thanks{These authors contributed equally to this work.}
\affiliation{CAS Key Laboratory of Quantum Information, University of Science and Technology of China, \\ Hefei 230026, People's Republic of China}
\affiliation{CAS Center For Excellence in Quantum Information and Quantum Physics, University of Science and Technology of China, Hefei, 230026, People's Republic of China}
\author{Manfredi Scalici}
\affiliation{Centre for Quantum Optical Technologies, Centre of New Technologies, University of Warsaw, Banacha 2c, 02-097 Warsaw, Poland}
\author{Jan Kołodyński}
\affiliation{Centre for Quantum Optical Technologies, Centre of New Technologies, University of Warsaw, Banacha 2c, 02-097 Warsaw, Poland}

\author{Guo-Yong Xiang}
\email{gyxiang@ustc.edu.cn}
\affiliation{CAS Key Laboratory of Quantum Information, University of Science and Technology of China, \\ Hefei 230026, People's Republic of China}
\affiliation{CAS Center For Excellence in Quantum Information and Quantum Physics, University of Science and Technology of China, Hefei, 230026, People's Republic of China}

\author{Chuan-Feng Li}

\affiliation{CAS Key Laboratory of Quantum Information, University of Science and Technology of China, \\ Hefei 230026, People's Republic of China}
\affiliation{CAS Center For Excellence in Quantum Information and Quantum Physics, University of Science and Technology of China, Hefei, 230026, People's Republic of China}

\author{Guang-Can Guo}

\affiliation{CAS Key Laboratory of Quantum Information, University of Science and Technology of China, \\ Hefei 230026, People's Republic of China}
\affiliation{CAS Center For Excellence in Quantum Information and Quantum Physics, University of Science and Technology of China, Hefei, 230026, People's Republic of China}

\author{Alexander Streltsov}
\email{a.streltsov@cent.uw.edu.pl}
\affiliation{Centre for Quantum Optical Technologies, Centre of New Technologies,
University of Warsaw, Banacha 2c, 02-097 Warsaw, Poland}

\begin{abstract}
Open quantum systems exhibit a rich phenomenology, in comparison to closed quantum systems that evolve unitarily according to the Schr\"odinger equation. The dynamics of an open quantum system are typically classified into Markovian and non-Markovian, depending on whether the dynamics can be decomposed into valid quantum operations at any time scale. Since Markovian evolutions are easier to simulate, compared to non-Markovian dynamics, it is reasonable to assume that non-Markovianity can be employed for useful quantum-technological applications. Here, we demonstrate the usefulness of non-Markovianity for preserving correlations and coherence in quantum systems. For this, we consider a broad class of qubit evolutions, having a decoherence matrix separated from zero for large times. While any such Markovian evolution leads to an exponential loss of correlations, non-Markovianity can help to preserve correlations even in the limit $t \rightarrow \infty$. For covariant qubit evolutions, we also show that non-Markovianity can be used to preserve quantum coherence at all times, which is an important resource for quantum metrology. We explicitly demonstrate this effect experimentally with linear optics, by implementing the required evolution that is non-Markovian at all times.
\end{abstract}

\maketitle

\emph{Introduction.}---In quantum resource theories~\cite{ChitambarRevModPhys.91.025001} correlations, such as entanglement~\cite{HorodeckiRevModPhys.81.865}, are seen as expendable resource to perform certain tasks, e.g. quantum teleportation~\cite{BennettPhysRevLett.70.1895}. On the other hand, every quantum setup we try to control is subject to noise, as it interacts with an environment~\cite{Breuer2002}. Historically, the way to treat these interactions and solve the equations of motion of the system, was with the aim of the Born-Markov approximation which assumes that the characteristic time evolution of the environment is very short with respect to the one of the system \cite{Manzano_2020}. In other words, the environment immediately loses memory of its contact with the system and is restored to its initial condition instantaneously. Over time, several different mathematical descriptions of this feature have been proposed \cite{RivasPhysRevLett.105.050403,ChruscinskiPhysRevLett.112.120404,BaePhysRevLett.117.050403,PollockPhysRevLett.120.040405,MilzPhysRevLett.123.040401,Rivas_2014}.

The description that we will adopt in this work is based on the notion of divisibility of a dynamical map.  An evolution is termed \emph{Markovian}, or CP-divisible, if it can be decomposed into~\cite{RivasPhysRevLett.105.050403,ChruscinskiPhysRevLett.112.120404,BaePhysRevLett.117.050403}: 
\begin{equation}
    \Lambda_t  = V_{t,s} \circ \Lambda_s, \label{eq:MarkovianDecomposition}
\end{equation}
where $V_{t,s}$ is a valid quantum operation for all \mbox{$t \geq s \geq 0$}. Note that this definition is essential, if one is to describe an open system dynamics without an explicit model of the environment. However, using the Stinespring dilation, it is possible to simulate any Markovian evolution by letting the system interact repeatedly and for a short period of time  with an ancilla, which we can control, and then reset it to its initial state. This is relevant e.g. for experiments, where one is able to use noisy ancillary systems. Since the interaction period can be made arbitrary small, any Markovian evolution can be simulated even if the ancilla system decoheres very quickly. Generally, for any differentiable Markovian evolution, the state evolves according to the (time-dependent) Gorini–Kossakowski–Sudarshan–Lindblad equation
\cite{KOSSAKOWSKI1972247,Gorini_Completely_1976,lindblad1976generators, OpenQuantBook, chruscinski2012markovianity}
\begin{align}
\frac{\mathrm{d}\rho(t)}{\mathrm{d}t} &= \mathcal{L}_{t} (\rho) = -i\left[H(t),\rho(t)\right] \label{eq:MarkovianLindblad}\\
&+ \sum_{i,j}\gamma_{ij}(t)\left(A_{i}\rho(t)A_{j}^{\dagger}-\frac{1}{2}\left\{ A_{j}^{\dagger}A_{i},\rho(t)\right\} \right), \nonumber
\end{align}
where $H(t)$ is a time-dependent Hermitian operator, and $\gamma_{ij}(t)$ are elements of a positive semidefinite matrix $\gamma(t)$, which we call \emph{decoherence matrix} \cite{hall2014canonical}.

Quantum dynamics which do not admit Eq.~(\ref{eq:MarkovianDecomposition}) are called \emph{non-Markovian}. They exhibit memory effects that manifest themselves via backflow of operationally relevant quantities from the environment to the system \cite{Rivas_2014}. In contrast to Markovian evolutions, the simulation of a non-Markovian dynamics requires to establish and control correlations between the system and an ancilla for a finite time \cite{Chiuri_2012, PhysRevResearch.2.023133}. Since Markovian evolutions are easier to simulate, it is reasonable to assume that they are less useful for some tasks, when compared to non-Markovian dynamics. Examples for tasks demonstrating the usefullness of non-Markovianity are swapping the sign of entropy production rate and preserving purity in the context of thermal operations \cite{Bhattacharya_2020}, and improving the fidelity of quantum teleportation under a noisy channel \cite{laine2014nonlocal}.

In this work, we explore the usefulness of non-Markovianity for preserving correlations in quantum systems. At first glance, the property of a dynamics to be Markovian or not does not seem to be related to its ability to preserve correlations. Both Markovian and non-Markovian evolutions can preserve entanglement and other types of correlations for all times, including the limit $t \rightarrow \infty$.  While any Markovian evolution leads to a monotonic decrease of entanglement~\cite{RivasPhysRevLett.105.050403,KolodynskiPhysRevA.101.020303} and mutual information~\cite{LuoPhysRevA.86.044101}, correlations can still survive for large times if $\gamma(t)$ vanishes fast enough, and the dynamics becomes asymptotically noiseless. Similarly, non-Markovian evolutions can also preserve or destroy correlations, or even lead to their periodic loss and recovery. 

Here, we show that the ability of an evolution to preserve correlations is still closely related to (non)-Markovianity. For this, we consider a very general class of evolutions, having the property that the eigenvalues of the matrix $\gamma$ are separated from zero for large times, so that the evolution does not become simply unitary at long time scales. We focus on qubit systems, which is enough to demonstrate the main features we are interested in. For this class of qubit dynamics, we show that any Markovian evolution leads to an exponential loss of correlations. These results suggest that correlations can only be preserved by using non-Markovianity. To make a fair comparison between Markovian and non-Markovian dynamics, we focus on covariant qubit evolutions. We show that the minimal loss of entanglement and mutual information occurs for eternally non-Markovian evolutions, i.e., the ones exhibiting non-Markovianity for all times $t>0$. While entanglement vanishes in the limit $t \rightarrow \infty$, the dynamics still preserves nonzero mutual information and quantum discord.

Since covariant evolutions exhibit symmetry with respect to a given Hamiltonian \cite{10.1007/BFb0106777}, its eigenbasis provides a natural reference for defining quantum coherence~\cite{BaumgratzPhysRevLett.113.140401,StreltsovRevModPhys.89.041003}. In case of two-level systems this corresponds to considering phase-covariant evolutions \cite{PhysRevLett.116.120801, filippov2020phase}, which cover all dynamics that respect rotational symmetry about an axis in the Bloch representation, e.g. the $z$-axis. In this case, we find the evolution which preserves quantum coherence for all finite times, including the limit $t \rightarrow \infty$. Interestingly, this dynamics converges to a map which has a 2-dimensional image, having finite coherence with respect to the reference basis (see Supplemental Material for more details). As quantum coherence is a resource useful for quantum metrology ~\cite{PhysRevA.94.052324}, this dynamics allows us to estimate a parameter $\omega$ encoded in the unitary $U = e^{-i\omega \sigma_z}$, leading to non-zero quantum Fisher information even in the limit $t \rightarrow \infty$.

Non-Markovianity has been experimentally demonstrated based on various platforms such as linear optics~\cite{liu2011experimental,orieux2015experimental,PhysRevA.101.052107,PhysRevA.98.053862,PhysRevA.90.050301,PhysRevA.83.064102,PhysRevA.84.032112,tang2015experimental}, nuclear magnetic resonance~\cite{PhysRevA.99.022107,bernardes2016high}, quantum dot~\cite{PhysRevLett.106.233601}, micromechanical system~\cite{groeblacher2015observation}, trapped ions~\cite{gessner2014local}, and superconducting qubits~\cite{PhysRevA.94.063848}. 
An attractive experimental platform for studying non-Markovian
effects is offered by photonic systems, where controlled interactions between different degrees of freedom, preparation of arbitrary quantum states, and a full state tomography are highly desirable and also appealing for testing fundamental paradigms of quantum
mechanics. Here, we experimentally demonstrate a quantum process, which is non-Markovian for all $t>0$, using an optical system, and observe the optimal preservation of quantum correlations.

\medskip

\emph{Markovian qubit evolutions destroy correlations.}---We now consider Markovian qubit dynamics, having the property that all eigenvalues of $\gamma$ are separated from zero, i.e., $\gamma(t) \geq c \openone$ for some $c>0$. We show that  such evolutions lead to the exponential decay of any kind of correlations.

\begin{proposition}
\label{prop:Correlations}
Let $\mathcal{L}_t$ be a Lindbladian giving rise to the qubit dynamics $\Lambda_t$. If there is a constant $c>0$ and time $T\geq 0$, such that $\gamma(t) \geq c \mathbbm{1}$ for all $t\geq T$, the corresponding qubit dynamics $\Lambda_t$ fulfills
\begin{equation}
    \min_{\sigma^A\otimes \sigma^B}\left\Vert\Lambda_t \otimes \id(\rho^{AB}) - \sigma^A\otimes \sigma^B\right\Vert_1 \leq 2 e^{-2ct}
\end{equation}
for all two-qubit states $\rho^{AB}$ and the trace norm $||M||_1 = \mathrm{Tr}\sqrt{M^\dagger M}$.
\end{proposition}
\noindent We refer to the Supplemental Material for the proof.

Proposition \ref{prop:Correlations} shows that certain Markovian qubit dynamics destroy all correlations in bipartite quantum states. Moreover, the decay of correlations happens exponentially fast. As we will see in the following, finely tuned non-Markovian systems can preserve certain correlations for all times, including the limit $t \rightarrow \infty$.

\medskip
\emph{Non-Markovian covariant evolutions preserve correlations and coherence.}---We now focus on non-Markovian quantum evolutions that could potentially exhibit slower rates of decay of entanglement and other quantum correlations. We will show that, apart from entanglement, non-Markovianity is useful for preserving coherence. As coherence is a basis-dependent quantity, we consider evolutions commuting with the unitary encoding the phase, which we assume to be in the $z$-direction. Hence, we restrict our discussion to \emph{covariant} evolutions \cite{PhysRevLett.116.120801, filippov2020phase}, with the following decoherence matrix in the Pauli basis:
\begin{equation} \label{eq:GammaCovariant}
    \gamma(t)=\begin{pmatrix}
    a(t) & -i  x(t) & 0 \\
    i x(t) & a(t) & 0 \\
    0 & 0 & f(t) \\
    \end{pmatrix}.
\end{equation}
The eigenvalues of $\gamma(t)$ are given by $a(t) \pm x(t)$, and $f(t)$. We refer to the Supplemental Material for further discussion of covariant qubit dynamics.

In the same spirit as in Proposition \ref{prop:Correlations}, we assume that all eigenvalues of $\gamma(t)$ are separated from zero for all $t > T$. If all eigenvalues become eventually positive, the evolution will become  Markovian and the correlations vanish, as described in Proposition~\ref{prop:Correlations}.
In the following, we will thus focus on the other case, where the matrix $\gamma(t)$ has negative eigenvalues. As we discuss in the Supplemental Material, the only negative eigenvalue of $\gamma(t)$ must be $f(t)$, as any other negative eigenvalue will not result in a valid quantum dynamics. 

We will now investigate the action of the time evolution on a two-qubit quantum state $\rho^{AB}$, focusing in particular on correlations in the system.
We consider a broad class of correlation quantifiers $\C$, making the only assumption that the amount of correlations does not increase under local noise:
\begin{equation}
    \C(\Phi \otimes \id [\rho^{AB}]) \leq \C(\rho^{AB}), \label{eq:Correlations}
\end{equation}
where $\Phi$ is an arbitrary local operation. In particular, Eq.~(\ref{eq:Correlations}) is true for the mutual information and any measure of entanglement~\cite{VedralPhysRevLett.78.2275,VedralPhysRevA.57.1619,HorodeckiRevModPhys.81.865}. Our goal in the following is to determine functions $f(t)$ leading to the minimal loss of correlations among all dynamics with given $a(t)$ and $x(t)$. More precisely, given a correlation quantifier $\C$, a two-qubit state $\rho^{AB}$, and time $t \geq 0$ we aim to maximize $\C(\Lambda_t \otimes \id [\rho^{AB}])$ over all functions $f(t)$. 

It is tempting to believe that the optimal solution for $f(t)$ will in general depend on the setup, in particular on the state and the correlation quantifier. Perhaps surprisingly, we will see in the following that the optimal choice of $f(t)$ is unique, giving rise to a quantum evolution which is non-Markovian for all $t>T$. 

\begin{proposition} \label{prop:Covariant}
For given functions $a(t)$ and $x(t)$ and time $T$ such that
$a(t) \geq |x(t)|$, for all $t>T$, the phase-covariant dynamics for which the loss of correlations is minimal
at any given time $t>T$
is given by the function $f(t)$ satisfying the equality
\begin{equation}
4 e^{-2 A(t) - 4 F(t)} + l_z(t)^2 = (1+e^{-2 A(t)})^2,  \label{eq:GammaOptimal}
\end{equation}
where $F(t) = \int_{0}^{t} f(\tau) \, d \tau$.
In particular, for $x(t)=0$, $f(t) = - a(t) \tanh A(t)$.
\end{proposition}
\noindent We refer to the Supplemental Material for the proof.

As an illustration, consider the phase-covariant dynamics $\Lambda_t$ for which $a(t)=a$, $x(t)=x$ are constants such that $a \geq |x|$.
The evolution of an initial qubit state $\rho_t = \Lambda_t \rho(\vec{r}_0) = \rho(\vec{r}_t)$ is given by
\begin{subequations} \label{eq:diaggamma}
\begin{align} 
r_{1,2}(t) &= \alpha(t) r_{1,2}(0), \\
r_3(t) &= \beta(t) r_3(0) - c(t).
\end{align}
\end{subequations}
with $\alpha(t)=e^{-a t-\int_0^t f(t)dt},\,\,
    \beta(t)=e^{-2 a t},\,\,
    c(t)=\frac{x}{a}(1-e^{-2 a t})$.Then, we can write the Choi-Jamio\l{}kowski~(CJ) state of this evolution as
\begin{align}
\label{eq:nonMarkovdiag}
    \Omega_t =\frac{1}{4}\begin{pmatrix}
    1+\beta(t) & 0 & 0 & 2\alpha(t) \\
    0 & 1-\beta(t) & 0 & 0\\
    0 & 0 & 1-\beta(t) & 0\\
    2\alpha(t) & 0 & 0 & 1+\beta(t)\\
    \end{pmatrix} \\
    - \frac{c(t)}{4} \, \textrm{diag}(1,-1,1,-1). \nonumber
\end{align}
For the resulting evolution $\Lambda_t$ to be completely positive,
we require~\cite{hall2008complete}: $4\alpha(t)^2+c(t)^2\leq(1+\beta(t))^2$. This inequality is saturated for all $t \geq 0$ if we choose $f(t)$ as in Eq.~(\ref{eq:GammaOptimal}). Note that in this case, $f(t)$ is negative for all $|x|\leq a$ and $t>0$.
It is straightforward to verify that the optimal choice of the function $f(t)$,
as in Proposition \ref{prop:Covariant}, is
\begin{equation}
    f(t) = - \frac{1}{2} a \left(1 - \frac{x^2}{a^2}\right)     \frac{\sinh 2 a t}{\cosh^2 a t - \frac{x^2}{a^2} \sinh^2 a t}
    \label{eq:GammaOptimalConst}
\end{equation}
(see also Prop.~4 in Ref.~\cite{filippov2020phase}).
In the special case when $x=0$ and the dynamics becomes unital, we have 
$f(t)=-a \tanh{a t}$.
This evolution (for $a=1$, up to a constant factor) was first proposed in \cite{hall2014canonical} (see 
Eq.\,(14) therein). In our work, this dynamics arises naturally as a family of evolutions which is optimal for preserving correlations. 

We will now consider implications of these results for concrete correlation quantifiers. We use entanglement negativity~\cite{ZyczkowskiPhysRevA.58.883,VidalPhysRevA.65.032314} as a measure of entanglement
\begin{equation}
    E(\rho)= \frac{||\rho^{T_B}||_1-1}{2},
\end{equation}
where $T_B$ denotes the partial transpose. We also consider the quantum mutual information $I(\rho) = S(\rho^A) + S(\rho^B) - S(\rho)$ with the von Neumann entropy $S(\rho) = -\tr(\rho \log_2 \rho)$.

For the optimal choice of $f(t)$ as in Eq.~\eqref{eq:GammaOptimalConst}, the negativity of the CJ state is given by 
\begin{equation}\label{eq:NeEvolution}
    E(\Omega_t) = \frac{1}{2}e^{-2 a t}.
\end{equation}
We see that the evolution $\Lambda_t$ preserves entanglement for all finite times, as the CJ state is entangled in this case. However, $\Lambda_t$ is entanglement breaking in the limit $t\rightarrow \infty$, as the CJ state becomes separable in this limit~\cite{Horodeckidoi:10.1142/S0129055X03001709}. 

Interestingly, the mutual information does not vanish in the limit $t\rightarrow \infty$:
 \begin{equation}\label{eq:MIevolution}
     \lim_{t \rightarrow \infty}I(\Omega_t)= \frac{h(p)}{2}
 \end{equation}
 with $p=\frac{1+\frac{x}{a}}{2}$ and the binary entropy $h(p)=-p\log_2{p}-(1-p)\log_2(1-p)$. Additionally, the CJ state exhibits a nonzero amount of quantum discord~\cite{PhysRevLett.88.017901,Henderson_2001}, a type of quantum correlations beyond entanglement. Quantum discord is useful for various quantum technological tasks~\cite{ModiRevModPhys.84.1655,Streltsov2014,Bera_2017}, an important example being distribution of entanglement between remote parties~\cite{StreltsovPhysRevLett.108.250501,ChuanPhysRevLett.109.070501,StreltsovPhysRevA.92.012335,Streltsov2017,FedrizziPhysRevLett.111.230504,VollmerPhysRevLett.111.230505,PeuntingerPhysRevLett.111.230506}. Following results in~\cite{PhysRevA.81.042105}, we obtain:
\begin{equation}\label{eq:disEvolution}
    \lim_{t \rightarrow \infty} Q(\Omega_t)=\frac{h\left(\frac{1+\frac{x}{a}}{2}\right)}{2} + h\left(\frac{1+\frac{1}{2}\sqrt{1-\left(\frac{x}{a}\right)^2}}{2}\right)-1,
\end{equation}
where $Q$ is quantum discord as defined in~\cite{PhysRevLett.88.017901,PhysRevA.81.042105}. 
In the case of $|x|<a$, the discord remains nonzero in the limit $t \rightarrow \infty$.  We refer to the Supplemental Material for more details.

As we will see in the following, non-Markovianity is also useful for preserving quantum coherence, a fact which can be used in quantum metrology. Since we consider covariant evolutions with dephasing matrix of the form~(\ref{eq:GammaCovariant}), coherence with respect to the eigenbasis of $\sigma_z$ is a meaningful quantity in this setup. A quantifier of coherence $C(\rho)$ vanishes for all states which are diagonal in the eigenbasis of $\sigma_z$, and moreover $C(\rho)$ is monotonic under incoherent operations~\cite{AbergQuant-ph/0612146,BaumgratzPhysRevLett.113.140401,StreltsovRevModPhys.89.041003}. These are quantum operations $\Lambda[\rho] = \sum_i K_i \rho K_i^\dagger$ having the property that each Kraus operator does not create coherence~\cite{BaumgratzPhysRevLett.113.140401,StreltsovRevModPhys.89.041003}.  Using similar arguments as in the proof of Proposition~\ref{prop:Covariant}, we can see that a covariant qubit evolution is optimal for preserving coherence at any time $t \geq 0$, if $f(t)$ satisfies  Eq.~(\ref{eq:GammaOptimal}). More details can be found in the Supplemental Material.

We now consider the $\ell_1$-norm of coherence, defined as $C_{\ell_1}(\rho) = \sum_{i\neq j}|\rho_{ij}|$~\cite{BaumgratzPhysRevLett.113.140401}. 
For a single-qubit state with Bloch vector $\boldsymbol{r} = (r_1,r_2,r_3)$, the $\ell_1$-norm of coherence reduces to $C_{\ell_1} = \sqrt{r_1^2 + r_2^2}$. Using Eqs.\,(\ref{eq:diaggamma}), we can evaluate $C_{\ell_1}$ as a function of time: 
\begin{equation}
    C_{\ell_1}(t) = e^{-a t - \int \limits_{0}^{t} f(\tau) d\tau } C_{\ell_1}(0), 
\end{equation}
where $C_{\ell_1}(0)$ is the initial amount of coherence at time $t=0$. The maximal amount of coherence at any time $t \geq 0$ is obtained for $f(t)$ given in Eq.~(\ref{eq:GammaOptimal}), leading to
\begin{equation}
\label{eq:coherencecoeff}
    C_{\ell_1}(t) = \frac{1}{2} C_{\ell_1}(0) \sqrt{ (1 + e^{-2 a t})^2 - \frac{x^2}{a^2}(1 - e^{-2 a t})^2} \,.
\end{equation}
Coherence in general does not vanish even in the limit $t\rightarrow \infty$, as long as $C_{\ell_1}(0) > 0$ and $|x| < a$.

Non-Markovianity is also useful in the context of quantum metrology~\cite{PhysRevA.82.042103}. Let us suppose a quantum state $\rho$ interacts with a device through the Hamiltonian
$H = \frac{\omega}{2} \sigma_{z}$.
We would like to estimate the value of the parameter $\omega$. We can use the fact that the evolution commutes with the Hamiltonian $H$ and, for a suitably chosen $f(t)$, preserves coherence in the basis $\{ |0\rangle, |1 \rangle \}$, 
to facilitate the estimation of $\omega$. The lower bound on the variance of the estimator of $\omega$ is given by the quantum Cramer-Rao bound \cite{PhysRevLett.72.3439}:
\begin{equation}
    (\Delta \omega)^2\geq\frac{1}{\mathcal{F}_{\omega}(\rho)},
\end{equation}
where $\mathcal{F}_{\omega}(\rho)$ is the quantum Fisher information.  The following closed formula is valid in the qubit case \cite{PhysRevA.87.022337}:
\begin{equation}
\label{eq:FishInf}
\mathcal{F}_{\omega}(\rho)=|\dot{\vec{r}}|^2+\frac{(\vec{r}\cdot\dot{\vec{r}})^2}{1-r^2}
\end{equation}
with $\vec{r}$, the Bloch vector and $\dot{\vec{r}} = \partial \vec r / \partial \omega$. In case of phase-covariant dynamics considered here, the second term always vanishes and $\dot{\vec{r}}=tC_{\ell_1}(t)(\cos{\omega t},-\sin{\omega t},0)$, leading to $\mathcal{F}_{\omega}(\rho) = t^2 C_{\ell_1}^2(t)$, with $C_{\ell_1}$ being the $\ell_1$-norm of coherence. Hence, the non-Markovian evolution that maximizes $C_{\ell_1}$ in Eq.~(\ref{eq:coherencecoeff}) also maximizes the quantum Fisher information~(\ref{eq:FishInf}).

\begin{figure*}[htp]
	\includegraphics[scale=0.08]{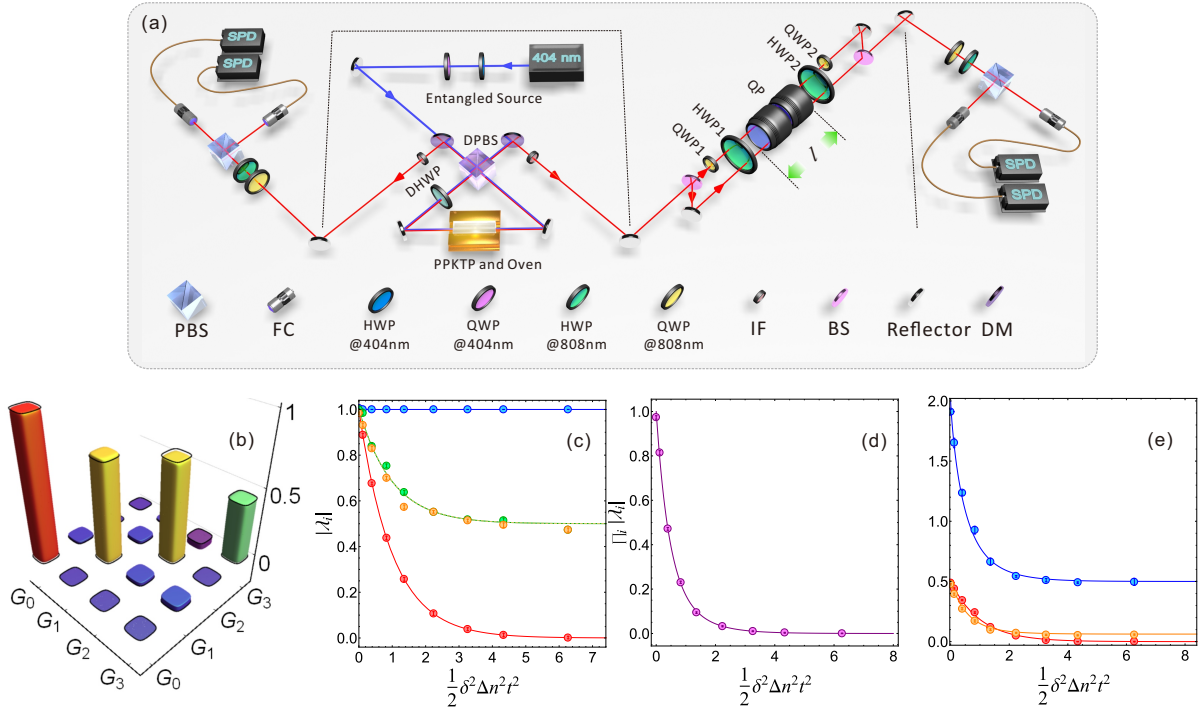}
	\caption{\label{fig:exp} 
	\textbf{Experimental setup for ENM process and results.} (a) The whole experimental setup includes three modules: Entangled photon source, eternally non-Markovian process, state tomography. (b) Experimentally reconstructed $F$ matrix with black-edged transparent cubes when $\frac{1}{2}\delta^2\Delta n^2t^2=0.91$. (c) The dynamical process of the absolutes of the spectral values of the process matrix, whose ideal values are given by Eq.~(\ref{eq:spectrum}) and are monotonic in time, in agreement with results in \cite{PhysRevLett.120.060406}. The dots are experimental results, and the lines are the corresponding theoretical fits. (d) The dynamical process of the product of the absolute values of the spectral values. (e) Dynamics of mutual information (light blue disks), negativity (red disks), and geometric discord (orange disks),  whose theoretical values are shown as solid lines. Key to components: PBS, polarizing beamsplitter; BS, beamsplitters; Q, quarter-wave plate; H, half-wave plate; SPD, single photon source; DHWP, dichroic half wave plate; DPBS, dichroic polarizing beamsplitter; DM, dichroic mirror; FC, fiber coupler.
	}
\end{figure*}

\medskip
\emph{Experimental implementation of eternally non-Markovian process.}---We now present optical experiments, demonstrating that non-Markovianity is useful for preserving quantum coherence and correlations, as predicted in the theoretical part of this work. We achieve the goal of simulating a non-Markovian evolution by utilizing the fact that it can be obtained as a mixture of different Markovian dynamics~\cite{megier2017eternal}. Several attempts of simulating non-Markovian dynamics have been reported. This includes studying the transition between weak (only non CP-divisible) and strong (non P-divisible) non-Markovianity~\cite{Bernardes2015}, experimental investigations to demonstrate the ambiguity of the extension of the definition of classical non-Markovianity to the quantum case~\cite{PhysRevA.97.020102, PhysRevLett.126.230401}, using the spectrum of an evolution over time to infer non P-divisibility~\cite{PhysRevLett.120.060406}, and  practical demonstration of the non-convex nature of Markovian and non-Markovian channels set~\cite{PhysRevA.101.052107}.

Our experimental setup is illustrated in Fig.~\ref{fig:exp} (a), which relies on three stages: state preparation, implementation of the non-Markovian evolution, and performing state tomography. Since the dynamics of interest is described by four $t$-parameterized Kraus operators, we utilize the frequency degree and path degree of one photon as the environment and the polarization of one of the photons as the system of interests, where the system-environment interaction is provided by the coupling between the frequency of the photons and the quartz crystal and path-dependent operations. We implement experimentally the eternally non-Markovian evolution which is optimal for preserving quantum correlations and quantum coherence, as predicted in Proposition~\ref{prop:Covariant}. Our  all-optical implementation shows a high fidelity, and the details are provided in the Supplemental Materials.

In order to verify that we implement the correct non-Markovian evolution, we perform process tomography and experimentally determine the spectrum of the corresponding $n^2\times n^2$ matrix $F$ which is derived from this dynamical process at each time, following \cite{PhysRevLett.120.060406}. In particular, each matrix element of $F$ in the basis of the corresponding Hilbert space can be determined as
\begin{equation}
F_{i,j}=\mathrm{Tr}[G_i\Lambda(G_j)],
\end{equation}
where $G_i=\sigma_i/\sqrt{2}$ and $\sigma_i$ are Pauli matrices. The critical experimental step in measuring the spectrum is the application of the dynamics to the basis matrices $G_{i\neq 0}$, which are not legitimate physical quantum states. Nevertheless, there always exists a finite real coefficient $c$ and two legitimate states $\rho_{i,1}$ and $\rho_{i,2}$ satisfying $G_{i\neq 0}=(\rho_{i,1}-\rho_{i,2})/c$, which makes the $F$ matrix and its eigenvalues $\{\lambda_i\}$ detectable in experiments.

In Figs.~\ref{fig:exp} (b-d), we present the results characterizing the spectrum of the process matrix for the relevant non-Markovian evolution. In particular, we compare it against the theoretical behavior of the process eigenvalues, whose moduli read (see Supplemental Material):
\begin{align}
&\{|\lambda_i|\}=1,\frac{1}{2}[1+\exp(-\frac{1}{2}\delta^2\Delta n^2 t^2)],\nonumber\\
&\frac{1}{2}[1+\exp(-\frac{1}{2}\delta^2\Delta n^2 t^2)],\exp(-\frac{1}{2}\delta^2\Delta n^2 t^2).\label{eq:spectrum}
\end{align}
Here, the environmental parameter $\delta$ corresponds to the variance of the frequency distribution and $\Delta n=n_H-n_V$ denotes the nonzero difference in the refraction indices of the $\ket{H}$ and $\ket{V}$ polarized photons. In particular, we verify that both the dynamics of each $|\lambda_i|$ as well as their product are in good agreement with the experimental data, which shows the high fidelity of our experimental implementation. In Fig.~\ref{fig:exp} (e), we show explicitly the resulting dynamics of entanglement negativity, quantum discord, and mutual information---indeed, the implemented non-Markovian evolution yields these three measures of correlations to follow the optimal behavior predicted in our work. 

\medskip
\emph{Conclusions.}---We have shown that non-Markovianity is useful for preserving correlations and coherence in quantum systems. Any Markovian qubit evolution leads to the exponential loss of correlations, if the decoherence matrix is separated from zero for large times. Non-Markovian qubit evolutions with this property can preserve mutual information and quantum discord for all times, including the limit $t \rightarrow \infty$. For covariant evolutions, we have shown that non-Markovianity is also useful for preserving quantum coherence with respect to the reference basis. This effect can be used for parameter estimation: a phase encoded in a covariant unitary can be estimated with finite precision at any time, and the quantum Fisher information is nonzero also in the limit $t \rightarrow \infty$. We characterize covariant qubit evolutions that are optimal for preserving quantum coherence and correlations, and implement them experimentally using linear optics. 

Our results suggest that if a certain degree of control over the noise is available, it may still be possible to distribute large amount of correlations over noisy channels. This is also demonstrated by our experiment, making our experimental methods applicable for studying fundamental problems in quantum information science. Non-Markovianity appears to be an important feature for quantum technologies, crucial to maintain and store information in the form of quantum correlations and superposition. 

We acknowledge financial support by the ``Quantum Optical Technologies'' project, carried out within the International Research Agendas programme of the Foundation for Polish Science co-financed by the European Union under the European Regional Development Fund and the "Quantum Coherence and Entanglement for Quantum Technology" project, carried out within the First Team programme of the Foundation for Polish Science co-financed by the European Union under the European Regional Development Fund.

\bibliographystyle{apsrev4-2}
\bibliography{refs}

\section*{Supplemental Material}

\subsection{Qubit Markovian dynamics}
Let us consider a two-level quantum system and its dynamical evolution given by a time-dependent Lindbladian:
\begin{equation}
\label{eq:Lindbladian}
    \mathcal{L}_{t} \rho =  \sum \limits_{i,j=1}^3 \gamma_{ij}(t) \left( 
    \sigma_i \rho \sigma_j -
    \frac{1}{2} \left \{\sigma_j \sigma_i , \rho \right \}
    \right),
\end{equation}
where $\{\sigma_i\}_{i=1,2,3}$ are Pauli matrices and the coefficients
$\gamma_{ij}(t)$ form a Hermitian matrix $\gamma(t) = (\gamma_{ij}(t))$, 
$\gamma(t) = \gamma(t)^{\dagger}$.
Eq.\,(\ref{eq:Lindbladian}) specifies the evolution of the system as an initial value problem:
\begin{equation}
\label{eq:initval}
    \frac{d}{dt} \rho(t) = \mathcal{L}_t \rho(t),  \quad \rho(0) = \rho_0.
\end{equation}
We will make use of the standard notation: $\dot{\rho}_t = \frac{d}{dt} \rho(t)$.

We assume that the decoherence matrix $\gamma(t)$ is such that  the solution to the above equation gives rise to a family $\Lambda_t$ of completely positive trace-preserving (CPTP) maps: $\rho(t) = \Lambda_t \rho_0$.
While it is difficult to obtain a general condition that $\gamma(t)$ must satisfy in order to generate a CPTP evolution,
several special cases have been considered in the literature
\cite{hall2008complete, filippov2020phase}.
Nevertheless, it is known that the condition $\gamma(t) \geq 0$ for all $t\geq0$
is necessary and sufficient for $\Lambda_t$ to be CP-divisible~\cite{chruscinski2012markovianity, Rivas2012}.

To reduce the problem of solving Eq.\,(\ref{eq:initval}) to a set of ordinary differential equations, where the quantum nature of the system is implicit in the choice of a suitable parametrisation, we use the notation: 
\begin{align}
    \rho(t) &= \frac{1}{2} \left(\mathbbm{1} + \sum_k a_k(t) \sigma_k\right),
\end{align}
$\vec{a}(t) = (a_1(t), a_2(t), a_3(t))$, 
$|| \vec{a}(t) || \leq 1$.
Using the commutation relations of the Pauli matrices, we obtain from Eq.\,(\ref{eq:Lindbladian}):
\begin{equation}
\label{eq:linsysidx}
    \mathcal{L}_t \rho(t) = \sum \limits_{i,k=1}^{3}
    \left(
    \frac{1}{2} (\gamma_{ik}(t) + \gamma_{ki}(t)) \, a_i(t) - \gamma_{ii}(t) \, a_k(t)) \sigma_k
    \right).
\end{equation}
Setting $\gamma(t)^S = \frac{1}{2}(\gamma(t) + \gamma(t)^T)$ and
$\vec{\xi}(t) = (\xi_k(t) )$, where $\xi_k(t) = i \sum_{i,j=1}^{3}\epsilon_{ijk} \gamma_{ij}(t)$,
we get
\begin{equation}
\label{eq:linsys}
    \dot{\vec{a}}(t) = (\gamma^S_t - (\tr\,\gamma(t)) \mathbbm{1}) \, \vec{a}(t) + \vec{\xi}(t),
    \quad
    \vec{a}(0) = \vec{a}_0.
\end{equation}

In the following, we will make the assumption that 
the matrix elements $\gamma_{ij}(t)$ are such that for all $i,j = 1,2,3$ and $0 \leq t_1 < t_2 < \infty$, the integrals 
$\int_{t_1}^{t_2} \mathfrak{Re} \gamma_{ij}(t) dt$,
and
$\int_{t_1}^{t_2} \mathfrak{Im} \gamma_{ij}(t) dt$ 
are finite.
We know from the general theory
(see Theorem 5.3 in \cite{hale1980ordinary}, p.30),
that in that case there exists a unique solution to Eq.\,(\ref{eq:linsys}) for $t \geq 0$.

A general solution to the inhomogeneous differential equation (\ref{eq:linsys}) is obtained in the usual way.  Let $X_t$ be the fundamental solution to the homogeneous equation: $\dot{\vec{a}}(t) = A_t \vec{a}(t)$,
where $A_t = \gamma^S_t - (\tr\,\gamma(t)) \mathbbm{1}$, i.e.
$\frac{d}{dt} X_t = A_t X_t$ and $X_0 = \mathbbm{1}$.
Then the solution to Eq.\,(\ref{eq:linsys}) is given by
\begin{equation}
    \label{eq:linsys_sol}
    \vec{a}(t) = X_t \vec{a}_0 + X_t \int \limits_{0}^{t} X_s^{-1} \vec{\xi}_s ds.
\end{equation}

As we can see from Eq.\,\eqref{eq:linsys_sol}, the evolution of a quantum two-level system given by $\mathcal{L}_t$ splits into a sum of two evolutions: one that represents a solution to the homogeneous system of ordinary differential equations (for which $\vec{\xi}_t = 0$, or equivalently, for which $\gamma(t)$ is a real symmetric matrix that generates a unital CPTP evolution) and the other that is independent of the initial condition of the system.

\subsection{Covariant qubit dynamics}
For any covariant qubit evolution, the decoherence matrix in the Pauli basis takes the form \cite{filippov2020phase}:
\begin{equation} \label{eq:GammaCovariant2}
    \gamma(t)=\begin{pmatrix}
    a(t) & -i  x(t) & 0 \\
    i x(t) & a(t) & 0 \\
    0 & 0 & f(t) \\
    \end{pmatrix}.
\end{equation}
Note that the Lindbladian, expressed in the basis $\{ \sigma_{+}, \sigma_{-}, \sigma_{3} \}$, where $\sigma_{\pm} = \frac{1}{2}(\sigma_{1} \pm \sigma_{2})$, has diagonal decorehence matrix $\gamma(t)$, satisfying the requirement for a general covariant quantum evolution \cite{10.1007/BFb0106777}.

For any covariant qubit dynamics the equations of motion~(\ref{eq:linsys_sol}) reduce to:
\begin{subequations}
\begin{align}
    \dot{r}_1(t) &= -[a(t) + f(t)] r_1(t), \\
    \dot{r}_2(t) &= -[a(t) + f(t)] r_2(t), \\
    \dot{r}_3(t) &= -2 a(t) r_3(t) - 2x(t). \label{eq:r3dot}
\end{align}
\end{subequations}

The solution to the above equations gives rise to a CPTP dynamics if and only if 
\begin{subequations}
\begin{align}
\label{eq:CPconditionA}
e^{-2 A(t)} + |l_z(t)| &\leq 1, \\
4 e^{-2 A(t) - 4 F(t)} + l_z(t)^2 &\leq (1+e^{-2 A(t)})^2 \label{eq:CPconditionB}
\end{align}
\end{subequations}
where
$A(t) = \int_{0}^{t} a(\tau) \, d \tau$,
$X(t) = \int_{0}^{t} x(\tau) \, d \tau$, 
$F(t) = \int_{0}^{t} f(\tau) \, d \tau$,
and $l_z(t) = 2 e^{-2 A(t)} \int_{0}^{t} x(\tau) \, e^{2 A(\tau)} \, d \tau$
(see Eqs.~(5) and (11) in Ref. \cite{filippov2020phase}).
We obtain
\begin{subequations}
\begin{align}
    r_1(t) &=  e^{ -A(t) - 2 F(t)} r_1(0), \\
    r_2(t) &=  e^{ -A(t) - 2 F(t)} r_2(0), \\
    r_3(t) &=  e^{ -2A(t)} r_3(0) + l_z(t).  \label{eq:r3solution}
\end{align}
\end{subequations}

We will now show that the only negative eigenvalue of $\gamma(t)$ must be $f(t)$, and any other negative eigenvalue will not result in a valid quantum dynamics. 
Let us assume by contradiction that one of two eigenvalues: $a(t) \pm x(t)$
is negative.  
Without loss of generality, we may say that there exists a constant $c>0$ such that $x(t) > a(t)+c$ for all $t > T$. From Eq.~(\ref{eq:r3dot}) we see that $\dot{r}_3(t) < -2c$ for all $t > T$, which could  not lead to a valid quantum evolution, as any Bloch vector would inevitable evolve into a vector outside of the Bloch ball. Thus, $\gamma(t)$ can have only one negative eigenvalue for all $t > T$, which must be $f(t)$.

\subsection{Proof of Proposition 1}
At first, let us assume that $T=0$.
Because $\gamma(t) \geq c \mathbbm{1}$, and hence $\gamma^S_t \geq c \mathbbm{1}$,
we can rewrite Eq.\,(\ref{eq:linsys}) as
\begin{equation}
    \dot{\vec{a}}(t) = (A'_t - 2c \mathbbm{1}) \,  \vec{a}(t) + \vec{\xi}(t),
\end{equation}
where 
$A'_t = \gamma^S_t  - \mathrm{tr} \gamma(t) \mathbbm{1} + 2c \mathbbm{1} < 0$.
The solution to the above equation can be written as
\begin{equation}
    \vec{a}(t) = e^{-2ct} X^{'}_t \vec{a}_0 + e^{-2ct} X^{'}_t \int \limits_{0}^{t} e^{2cs} (X^{'}_s)^{-1} \vec{\xi}_s ds.
\end{equation}
Here, $X^{'}_t$ represents a valid CPTP dynamics: $\frac{d}{dt} X^{'}_t = A^{'}_t X^{'}_t$ and $X^{'}_0 = \mathbbm{1}$.
If by $\vec{\eta}(t)$ we denote the vector
\begin{equation}
    \vec{\eta}(t) = e^{-2ct} X^{'}_t \int \limits_{0}^{t} e^{2cs} (X^{'}_s)^{-1} \vec{\xi}_s ds,
\end{equation}
then $| \vec{a}(t) - \vec{\eta}_t| \leq 2 e^{-2ct} |\vec{a}_0|$.
Hence 
\begin{equation}
    ||\Lambda_t \rho_0 - \tilde{\rho}(t)||_1 \leq e^{-2ct},
\end{equation}
where $\tilde{\rho}(t) = \frac{1}{2} (\mathbbm{1} + \vec{\eta}_t \cdot \vec{\sigma})$
for any state $\rho_0$.
This implies
\begin{equation}
    ||\Lambda_t - \Phi_t|| \leq e^{-2ct}, \label{eq:limit}
\end{equation}
where $\Phi_t \rho = (\tr \rho)\tilde{\rho}(t)$
and the norm of a linear map is given by the infimum over all quantum states
\begin{equation}
    ||\Lambda_t - \Phi_t|| = \inf \limits_{\rho} ||\Lambda_t \rho - \Phi_t \rho||_1 \label{eq:normMap}.
\end{equation}

Recall that the following inequality holds true for any pair of quantum channels $\Lambda_1$ and $\Lambda_2$ acting on a Hilbert space of dimension $d$ (see \cite{effros2000operator}, Corollary 2.2.4):
\begin{equation}
    || \Lambda_1 \otimes \mathbbm{1}_d - \Lambda_2 \otimes \mathbbm{1}_d|| \leq d||\Lambda_1 - \Lambda_2||,
\end{equation}

With Eq.~(\ref{eq:limit}), it follows that 
\begin{equation}
   || \Lambda_1 \otimes \mathbbm{1}_d - \Lambda_2 \otimes \mathbbm{1}_d|| \leq 2 e^{-2ct}.
\end{equation}
The action of $\Phi_t$ on one qubit of a two-qubit state $\rho^{AB}$ is 
\begin{equation}
    \Phi_t\otimes \id (\rho^{AB}) = \Phi_t(\rho^A) \otimes \rho^B.
\end{equation}
We obtain 
\begin{equation}
    ||\Lambda_t \otimes \id(\rho^{AB}) - \Phi_t(\rho^{A}) \otimes \rho^B||_1 \leq 2 e^{-2ct},
\end{equation}
for any two-qubit state $\rho^{AB}$.
Finally, if $T>0$,
we can repeat the argument above for the
evolution $\Lambda'_t = \Lambda_{t+T} \Lambda_{T}^{-1}$,
making use of the fact that $\Lambda_t$ is CP-divisible.
This completes the proof.

\subsection{Proof of Proposition 2}

Let $\gamma(t)$ be as in Eq.~\eqref{eq:GammaCovariant} of the main text.
Suppose $f_0(t)$ is the function that satisfies Eq.~\eqref{eq:GammaOptimal} of the main text.
We can write the decoherence matrix $\gamma(t)$ as a sum of two matrices:
\begin{equation} \label{eq:GammaCovariantSplit}
    \gamma(t)=\begin{pmatrix}
    a(t) & -i  x(t) & 0 \\
    i x(t) & a(t)  & 0 \\
    0 & 0 & f_0(t) \\
    \end{pmatrix} + 
    \begin{pmatrix}
    0 & 0 & 0 \\
    0 & 0 & 0 \\
    0 & 0 & f(t) - f_0(t) \\
    \end{pmatrix}.
\end{equation}
It is easy to see that as long as both matrices generate valid CPTP dynamics independently, the resulting evolutions commute.
Since $f_0(t)$ satisfies Eq.~\eqref{eq:GammaOptimal} of the main text, the first matrix generates a valid CPTP dynamics.
The dynamics generated be the second matrix is clearly CPTP for $t>T$.
Indeed, because for all $t>T$, $a(t) > 0$, and hence $A(t)>0$, 
then from Eq.~\eqref{eq:CPconditionB}, we have that 
$0 \leq F(t)-F_0(t) = \int_{0}^{t} (f(\tau) - f_0(\tau)) \, d \tau$.
This is enough to satisfy the complete-positivity conditions \eqref{eq:CPconditionA} and \eqref{eq:CPconditionB}.
According to Eq.~\eqref{eq:Correlations} of the main text, the amount of correlations at any given time $t>T$ cannot be larger than in the optimal case when $f(t) = f_0(t)$.

The second part of the Proposition follows immediately from observing that the function $l_z(t)$ vanishes as long as we put  $x(t) = 0$.

Using similar arguments, we can see that the optimal preservation of quantum coherence at any time $t \geq 0$ is achieved if $f(t)$ is chosen such as to satisfy  Eq.~(\ref{eq:GammaOptimal}) of the main text. 
Indeed, the map generated by the second matrix in Eq.~\eqref{eq:GammaCovariantSplit}
is a CPTP dynamics that does not create coherence (a phase-damping map) and hence the value of
any quantifier of coherence cannot increase under its action.

\subsection{Geometry of image states}

For the sake of simplicity, we show again the solution (Eq.~\eqref{eq:diaggamma} of the main text):
\begin{gather}
    r_{1,2}(t)=r_{1,2}(0)\alpha(t)\\
    r_{3}(t)=\beta(t)r_3(0)-c(t),
\end{gather}
with 
\begin{gather}
\alpha(t)=e^{-a t-\int_0^t f(t)dt}\\
    \beta(t)=e^{-2\gamma t}\\
    c(t)=\frac{x}{\gamma}(1-e^{-2\gamma t}).
\end{gather}
Choosing the optimal $f(t)$ defined in Eq.~(\ref{eq:GammaOptimalConst}) of the main text:
\begin{equation}
    \alpha(t)=\sqrt{\frac{(1+e^{-2\gamma t})^2}{4}-\left(\frac{x}{\gamma}\right)^2\frac{(1-e^{-2\gamma t})^2}{4}}
\end{equation}
Then it's clear that in the limit $t\rightarrow\infty$, the Bloch sphere becomes a flat disk of radius $\frac{1}{2}\sqrt{1-\left(\frac{x}{\gamma}\right)^2}$ with the center at $\frac{x}{\gamma}$ along the $z$-axis.

\subsection{Evaluation of Quantum Discord}
We follow the results in \cite{PhysRevA.81.042105} regarding  the quantum discord of 4x4 X-states. The classical part of the correlations is given by Eq.~(22) in \cite{PhysRevA.81.042105} and it involves the minimization of the conditional entropy (22) (conditional respect to general von Neumann measurements $B_i$). The entropies $S(\rho_0)$ and $S(\rho_1)$ are defined in Eqs.~(19) and (20) in \cite{PhysRevA.81.042105}  and the parameters $\theta$ and $\theta'$ in Eqs.~(16) and (17) in \cite{PhysRevA.81.042105}. In our case, (see Eq.~\eqref{eq:nonMarkovdiag} of the main text),
\begin{gather}
    \rho_{11}=\rho_{33}=\frac{1}{4}\left(1+\frac{x}{a}\right)\\
    \rho_{22}=\rho_{44}=\frac{1}{4}\left(1-\frac{x}{a}\right)\\
    \rho_{14}=\rho_{41}=\frac{1}{4}\sqrt{1-\left(\frac{x}{a}\right)^2}\\
    \rho_{23}=\rho_{32}=0\\
    \Theta=4kl\left(\frac{\sqrt{1-\left(\frac{x}{a}\right)^2}}{4}\right)^2\\
    \theta=\sqrt{\frac{4kl\left(\frac{\sqrt{1-\left(\frac{x}{a}\right)^2}}{4}\right)^2}{\left[\frac{1}{2}\left(1+\frac{x}{a}\right)k+\frac{1}{2}\left(1-\frac{x}{a}\right)l\right]^2}}\\
    \theta'=\sqrt{\frac{4kl\left(\frac{\sqrt{1-\left(\frac{x}{a}\right)^2}}{4}\right)^2}{\left[\frac{1}{2}\left(1+\frac{x}{a}\right)l+\frac{1}{2}\left(1-\frac{x}{a}\right)k\right]^2}},
\end{gather}
where $k$ and $l$ are the parameters of the measurements $\{B_i\}$. The minimum of the conditional entropy is attained in one of the three cases:
\begin{itemize}
    \item $k=0$, $l=1$
    \item $k=1$, $l=0$
    \item $k=l=\frac{1}{2}$.
\end{itemize}
In  the first two cases, $\theta=\theta'=0$ and $S(\rho_0)=S(\rho_1)=1$, which is not the minimal value. In the third case, $\theta=\theta'=\frac{\sqrt{1-\left(\frac{x}{a}\right)^2}}{2}=\theta_{max}$. 
Because the reduced state $\rho^A$ is the maximally mixed one, we obtain
\begin{equation}
\label{ClassCorr}
    \mathcal{C}(\rho_X)=1-S(\rho_0)|_{\theta_{max}},
\end{equation}
where $\mathcal C$ is the measure of classical correlations as defined in~\cite{Henderson_2001,PhysRevA.81.042105}. Then the quantum discord is computed as the difference between the total correlations, given by the mutual information in Eq. (\ref{eq:MIevolution}), and Eq. (\ref{ClassCorr}).

\subsection{Experimental details}
The whole experimental set up is shown in Fig.~\ref{fig:exp} of the main text, it consists of three parts: state preparation, eternally non-Markovian (ENM) process, and state tomography.

In the state preparation module, we can experimentally generate arbitrary pure qubit states 
\begin{equation}
\ket{\phi}=\cos(\alpha)\ket{H}+e^{-i\beta}\sin(\alpha)\ket{V}.
\end{equation}

In the ENM process module, we can experimentally implement the process with probability $1/2$, as shown in Fig.~\ref{fig:exp} of the main text. In particular, assume that we have an arbitrary qubit states (in basis $\{\ket{H},\ket{V}\}$)
\begin{align}
\rho_0=\frac{1}{2}\left(I+x_0\sigma_x+y_0\sigma_y+z_0\sigma_z\right).
\end{align}
The first 50:50 beam splitters (BS) separate the photons into approximately two branches with equal probabilities independent of the polarization of the photons. 

The upper branch is reflected by a mirror, and passes through a half-wave plate (HWP) with angle $22.5$, implementing the unitary operation
\begin{align}
u_1=h(22.5)=\frac{1}{\sqrt{2}}\left(\sigma_x+\sigma_z\right).
\end{align}  
The lower branch goes through a quarter wave plate (QWP) with angle $0$, followed by $h(22.5)$, resulting in the transformation
\begin{align}
u_2=h(22.5)q(0)=\frac{1}{\sqrt{2}}\begin{pmatrix}
&1&1\\&1&-1
\end{pmatrix}\begin{pmatrix}
&1&0\\&0&i
\end{pmatrix}=\frac{1}{\sqrt{2}}\begin{pmatrix}
&1&i\\&1&-i
\end{pmatrix}.
\end{align}
The overall state then becomes
\begin{align}
\rho_1=\frac{1}{2}u_1\rho_0u_1^\dag+\frac{1}{2}u_2\rho_0u_2^\dag.
\end{align}

After these two wave plates, the overall state goes through a decoherence process in birefringent crystal. This is an open quantum system dynamics, where the open system is provided by the polarization of the single photons and the environment is provided by the frequency of the photons. 

Similar to the previous works~\cite{liu2011experimental,wu2020detecting}, we make use of a FP cavity to modify the spectrum of the photons, resulting in a non-Guassian profile, which will lead to a non-Markovian process. If the spectrum of the frequency can be approximately modeled by Guassian profile, then the decoherence between $\ket{H}$ and $\ket{V}$ photons is Markovian. 

In this experiment, we will not modify the spectrum of the photons, which means we will not change the spectrum of the frequency of the single photons. The decoherence between $\ket{H}$ and $\ket{V}$ can be modeled by a unitary evolution
\begin{align}\label{unitaryt}
U_{tot}(t)=&\int\mathrm{d}\omega[\,\exp(-\mathrm{i}n_H\omega t)\ketbra{\omega}{\omega}\otimes\ketbra{H}{H}\nonumber\\&+\exp(-\mathrm{i}n_H\omega t)\ketbra{\omega}{\omega}\otimes\ketbra{V}{V}\,],
\end{align}
and our environment is chosen as
\begin{equation}
    \ket{\phi_E}=\int\mathrm{d}\omega f(\omega)\ket{\omega}.
\end{equation}

The corresponding reduced dynamical map $\Lambda_t$ of the polarization degrees of freedoms takes the form,
\begin{subequations}
	\begin{align}
	&\ketbra{H}{H}\xrightarrow{\Lambda_t}\ketbra{H}{H},\\
	&\ketbra{V}{V}\xrightarrow{\Lambda_t}\ketbra{V}{V},\\
	&\ketbra{H}{V}\xrightarrow{\Lambda_t} \kappa(t)\ketbra{H}{V},\\
	&\ketbra{V}{H}\xrightarrow{\Lambda_t} \kappa^*(t)\ketbra{V}{H},
	\end{align}
\end{subequations}
where the decoherence factor reads
\begin{equation}
\kappa(t)=\int\mathrm{d}\omega |f(\omega)|^2\exp(-\mathrm{i}\Delta n\omega t),
\end{equation}
and $\Delta n=n_H-n_V\approx0.0089$ denotes the nonzero difference in the refraction indices of the $\ket{H}$ and $\ket{V}$ polarized photons. 

The spectral of the single photons $|f(\omega)|^2$ in our experiments admits a Guassian distribution, i.e., 
\begin{equation}
|f(\omega)|^2=\frac{1}{\sqrt{2\pi}\delta}\exp\left[-\frac{(\omega-\omega_0)^2}{2\delta^2}\right],
\end{equation}
where $\omega_0$ is the central frequency and $\delta\approx 1.44\times 10^{12}$Hz is the variance, corresponding to the linewidth $\Delta\lambda\approx 0.5$nm of down converted photons~\cite{Fedrizzi:07}. One can check that the normalization holds, i.e.,  $\int\mathrm{d}\omega |f(\omega)|^2=1$.

Then the decoherence factor decays exponentially with $t^2$, or equivalently the square of the crystal length $l^2$. We can calculate the decoherence factor and it can be written as 
\begin{align}
    \kappa(l)=\exp\left(-\frac{\Delta n^2\delta^2l^2}{2c^2}-\frac{i\Delta n\omega_0l}{c}\right),
\end{align}
where $l$ is the length of the crystal, $c$ is the velocity of light. We can check that $|\kappa(l)|=\exp\left(-\frac{\Delta n^2\delta^2l^2}{2c^2}\right)$ decays exponentially according to $l^2$. In our experiment, we extract the value of $\frac{1}{2}\delta^2\Delta n^2t^2$ for each quartz plate from the data of process tomography, instead of estimating $\kappa(l)$ with empirical formula.

Thus in principle we can implement the above process based on the optical setup in Fig.~\ref{fig:exp} (b) of the main text. However, due to the divergence of the optical path in the birefringence crystal for different locations of the cross section, there is an unpredictable phase $\phi_i$ between $H$ and $V$ polarized photons in each path. These phases are not equal to $\Delta n\omega_0l/c$ due to an imperfect fabrication of the birefringence crystal. To eliminate 
this divergence, we need to tune the phase in the two paths separately. In particular we can place a phase compensator (PC, e.g., non-true-zero-order wave plates) crystal in each path to remove the phase, which results in the actual setup in Fig.~\ref{fig:exp} (c) of the main text.
 
After the decoherence process, photons in the upper branch are in the state
\begin{equation}
    \rho_u=\frac{1}{2}\left[I+x_0\sigma_z+|\kappa(l)|(z_0\sigma_x-y_0\sigma_y)\right],
\end{equation}
and the state in the lower branch is
\begin{equation}
    \rho_l=\frac{1}{2}\left[I-y_0\sigma_z+|\kappa(l)|(z_0\sigma_x-x_0\sigma_y)\right],
\end{equation}
then the upper branch passes through $h(22.5)$ and is converted to
\begin{equation}
    \rho_u'=\frac{1}{2}\left[I+|\kappa(l)|z_0\sigma_z+x_0\sigma_x+|\kappa(l)|y_0\sigma_y\right],
\end{equation}
while the photons in the lower branch are transformed to 
\begin{equation}
    \rho_l'=\frac{1}{2}\left[I+|\kappa(l)|z_0\sigma_z+|\kappa(l)|x_0\sigma_x+y_0\sigma_y\right].
\end{equation}

The final BS and mirror recombines the two branches and the final state is 
\begin{equation}
  \rho=\frac{1}{2}\left[I+\kappa(l)z_0\sigma_z\right]+\frac{1}{4}\left[1+\kappa(l)\right](x_0\sigma_x+y_0\sigma_y),
\end{equation}
thus we can realize the ENM process with $l$ corresponding to the length of the crystal.

Experimentally, the dynamical behavior of relevant physical quantities can be estimated from the reconstructed density matrix for each evolution time $t$. For an experimentally reconstructed state $\rho_t$, the negativity $E$, mutual information $I$, and geometric discord $D$ can be evaluated directly using $E(\rho_t)=\frac{\|\rho^{T_B}_t\|_1-1}{2}$, $I(\rho_t)=S(\rho^A_t)+S(\rho^B_t)-S(\rho)$, and $D(\rho_t)=\frac{1}{4}(\|\mathbf{x}\|^2+\|\mathbf{T}\|^2-\lambda_{max})$, where $x_i=\tr(\sigma_i\otimes\mathbb{I})\rho_t$, $T_{ij}=\tr(\sigma_i \otimes \sigma_j)\rho_t$, and $\lambda_{\max}$ is the largest eigenvalue of the matrix $K = \boldsymbol{x}\boldsymbol{x}^T + TT^T$~\cite{DakicPhysRevLett.105.190502}.

\end{document}